\documentclass[12pt]{article}
\usepackage[pctex32]{graphics}
\begin{document}
\begin{center}
  
{\Large \bf On the No-Go Theorem of  Supersymmetry Breaking}

\vspace{1cm}

                      Wung-Hong Huang*\\
                       Department of Physics\\
                       National Cheng Kung University\\
                       Tainan,70101,Taiwan\\

\end{center}
\vspace{2cm}

{\begin{center} {\large \bf  ABSTRACT} \end{center}}

    It is proved that, even if the gauge symmetry has been broken spontaneously at tree level, supersymmetry would never break through any finite orders of perturbation if it is not broken classically.

\vspace{4cm}
\begin{flushleft}
*E-mail:  whhwung@mail.ncku.edu.tw\\
\end{flushleft}
Physics Letters B179 (1986) 92

\newpage

    Supersymmetry [1 ], which is the only graded Lie  algebra of symmetries of the S-matrix that is consistent with the relativistic quantum field theory [2],  could exhibit improved ultraviolet behavior [3,4] and  provide us with a solution of the gauge hierarchy  problem [5]. It has been shown that, in a broad class  of field theories [6], if supersymmetry is not broken at the classical level then it is not broken by radiative corrections. Those are the models with only chiral superfields and models whose gauge symmetry and supersymmetry are unbroken classically.

   Recently Ovrut and Wess [7] have examined a class of supersymmetric theories whose internal symmetry is completely broken spontaneously. After extending the $R_{\zeta}$ gauge-fixing [8] procedure to the supersymmetric theories they calculated the one-loop corrections to the auxiliary fields. It is found that the invariance of the superpotential under the complexification of the internal symmetry group allows the vacuum expectation values of scalar fields to be adjusted to make the corrected D terms vanish while keeping the F terms zero. Hence, the supersymmetry is not broken.  In this paper, without explicitly calculating the quantum correction, we will prove that this another no-go theorem is always true in any supersymmetric gauge theory.  Furthermore, the internal symmetry is allowed not to be spontaneously broken completely.

   To break the supersymmetry in finite orders of perturbation, one must get an expectation value of the F or D field. Using the supergraphic techniques
[4] it can be easily seen that the F field does not give an expectation value if supersymmetry is unbroken at the classical level. Also, there is no induced expectation value of the D term if both gauge symmetry  and supersymmetry are unbroken classically [9]. In other words, the Coleman-Weinberg mechanism [10]
is not possible. This is a no-go theorem that has been well established. When the gauge symmetry is already  broken spontaneously at the classical level, quantum
corrections for the D terms (but not F terms) are induced. In the following sections we shall show that, for any induced D field, through some suitable transformations of complexification of the internal symmetry group one can always find vacuum expectation values of the scalar fields to make the D terms vanish.  (Recall that the superpotential is invariant under the complexified internal symmetry group so that the F terms remain zero.) Hence, supersymmetry is not broken as claimed by Ovrut and Wess [7].

    Consider first the supersymmetric theory with abelian gauge symmetry group. Let $\Phi_{+i}$,$\Phi_{-j}$ , and $\Phi_{0k}$ be chiral superfields with positive charge $q_{+i}$, negative charge $q_{-j}$ and zero charge, respectively. In order to preserve the supersymmetry in the n-th order of perturbation, there must exist the vacuum expectation values $a_{+i}^{(n)}$, $a_{-j}^{(n)}$ and $a_{0k}^{(n)}$ of scalar fields in the chiral superfields, $\Phi_{+i}$, $\Phi_{-j}$ , and $\Phi_{0k}$, which satisfy

$$\sum_i q_{+i} |a_{+i}^{(n)}|^2 + \sum_j q_{-j} |a_{-j}^{(n)}|^2 = C^{(n)}(a_{+i}^{(n)}, a_{-j}^{(n)}, a_{0k}^{(n)}), \eqno{(1)}$$
where the functional form of  $C^{(n)}$ depends on the order of perturbation. At tree level, the  $C^{(n)}$ term does not show up. After the quantum correction, it is induced. To find the solution of (1) we set, in the spirit of perturbation theory, the values of $a_{+i}^{(n)}$, $a_{-j}^{(n)}$, and $ a_{0k}^{(n)}$ in the argument of the function $C^{(n)}$ to the $(n-1)$th order only and the equation which we need to solve becomes

$$\sum_i q_{+i} |a_{+i}^{(n)}|^2 + \sum_j q_{-j} |a_{-j}^{(n)}|^2 = C^{(n)}( a_{+i}^{(n-1)}, a_{-j}^{(n-1)}, a_{0k}^{(n-1)}), \eqno{(2)}$$
\\
The existence of a solution for the above equation can be proved by the following observations. 

     First, when the gauge group is already broken classically then there must exist at least one nonzero vacuum expectation value among, $a_{+i}^{(0)}$ or $a_{-j}^{(0)}$ which satisfies

$$\sum_i q_{+i} |a_{+i}^{(0)}|^2 + \sum_j q_{-j} |a_{-j}^{(0)}|^2 = 0. \eqno{(3)}$$
Furthermore, because $q_{+i}$ is positive while $q_{-j}$ is negative, there must exist at least one nonzero value $a_{+i}^{(0)}$ accompanied with at least one nonzero value $a_{+i}^{(0)}$ in order to satisfy the above equation.

    Next, consider the one-loop correction. We see that the $C^{(1)}$ term is only a function of tree-level vacuum expectation values, so it can be regarded as a fixed number when we want to find the one-loop vacuum expectation values $a_{+i}^{(1)}$, $a_{-j}^{(1)}$, $a_{0k}^{(1)}$ which satisfy eq.(2). Therefore, regardless of the value of $C^{(1)}$, it is easily seen that we can always find a real parameter $\theta$ such that $a_{+i}^{(1)}$ and  $a_{-j}^{(1)}$, which we are searching now, are just the transformed values of $a_{+i}^{(0)}$ and  $a_{-j}^{(0)}$ under the complexified internal symmetry group, i.e.

$$a_{+i}^{(1)}=a_{+i}^{(0)} exp(q_{+i}\theta), ~~~a_{-j}^{(1)}=a_{-j}^{(0)} exp(q_{-j}\theta).    \eqno{(4)}$$
For example, if $C^{(1)}$ is positive (negative) then $\theta$ must be positive (negative).The point is that there exists at least one nonzero value $a_{+i}^{(0)}$ accompanied with at 1east one nonzero value $a_{-j}^{(0)}$. For higher loop corrections, the above discussion can still be used.

   We next examine supersymmetric theories with non-abelian internal symmetry which is broken spontaneously.  But the complete breaking is not necessary.
Let $T^{\alpha}$ be the generators of the gauge group and the chiral superfield $\Phi$ be its representation. (Extension to many different representations is straightforward.)  The vacuum expectation value of the scalar field in $\Phi$ is denoted as $a$. To preserve the supersymmetry in the $n$th order of perturbation there must exist an $a^{(n)}$  to satisfy 

 $$ {a^{(n)}}^{\dag} T^{\alpha} a^{(n)} = C^{(n)\alpha} (a^{(n)}),               \eqno{(5)}$$
where the functional form of $C^{(n)\alpha}$ depends on the order of the perturbation. In the spirit of perturbation theory, as in the abelian case, we let $a^{(n)}$ in the argument of $C^{(n)\alpha}$ be $a^{(n-l)}$ and (5) becomes

$${a^{(n)}}^{\dag} T^{\alpha} a^{(n)} =C^{(n)\alpha} (a^{(n-1)}),               \eqno{(6)}$$

     To solve $a^{(n)}$ in the above equation we can regard $a^{(n-l)}$ as fixed numbers. Using this property, we prove that there always exists a solution $a^{(n)}$ which satisfies (6) regardless of the value of $C^{(n)\alpha}$ which depends on the model and the order of perturbation.  Hence, supersymmetry is not broken through quantum corrections.

    Consider a system whose gauge symmetry has been spontaneously broken at tree level; there exists at  least a generator $T^r$ which satisfies

 $$T^r a^{(0)} \not=  0 ,       \eqno{(7)}$$
\\
where $T^r$ belongs to the Cartan subalgebra with diagonal elements only, i.e.

 $$T^r_{ij} = t_i^r \delta_{ij}.      \eqno{(8)}$$
\\
 Eq. (7) then tells us that there must exist at least one  nonzero value  $t_i^r a_i^{(0)}$.  Also, eq. (6) becomes

$${a^{(0)}}^{\dag} T^r a^{(0)} =\sum_i t_i^r |a_i^{(0)}|^2 = 0.   \eqno{(9)}$$
\\
 This shows that there must exist at least one positive and one negative value among $t_i^r |a_i^{(0)}|^2$.

    If we rotate $a^{(0)}$ in Cartan subspace of the complexified internal symmetry group

$$a^{(0)}\rightarrow a^{(0)}exp({1\over2}\sum_r T^r \theta^r )\equiv {a^{(0)}}' ,     \eqno{(10)}$$
and define 

$$L(\theta^r) \equiv {|{a^{(0)}}'|}^2 = \sum_i  {|a_i^{(0)}|}^2 exp\left({1\over2} \sum_r t_i^r \theta^r \right),   \eqno{(11)}$$
\\
then it is obvious that the slopes of $L(\theta^r)$ are just ${a^{(0)}}'^{\dag} T^r {a^{(0)}}'$ and can take on any value. It is this property that, no matter what the value of $\sum_{\alpha}|C^{(n)\alpha}|^2$, we can always find a set of $\theta^r$ to satisfy

      $$\sum_{\alpha}|{a^{(0)}}'^{\dag} T^r {a^{(0)}}'|^2 = \sum_{\alpha}|{C^{(n)\alpha}}|^2. \eqno{12)}$$
 
   Finally, with the fact that the ${a^{(0)}}'^{\dag} T^r {a^{(0)}}'$ are transformed as vectors in the regular representation of the internal symmetry group, we can rotate them to satisfy (6). Therefore, the solution of (6) is found and supersymmetry is still preserved at the quantum level.

\newpage
{\begin{center} {\large \bf  REFERENCES} \end{center}}
\begin{enumerate}
\item J. Wess and B. Zumino, Nucl. Phys. B 70 (1974) 39.
\item  R. Hagg, J. Lopuszanski and M. Sohnius, Nucl. Phys. B 88 (1975) 257.
\item  J. Wess and B. Zumino, Phys. Lett. B 49 (1974) 52;\\
         J. Iliopoulos and B. Zumino, Nucl. Phys. B 79 (1974) 310.
\item  M.T. Grisaru, W. Siegel and M. Rocek, Nucl. Phys. B 159 (1979) 429.
\item  E. Witten, Phys. Lett. B 105 (1981) 267.
\item B. Zumino, Nucl. Phys. B 89 (1975) 535;\\
        W. Lang, Nucl. Phys. B 114 (1976) 123;\\
        S. Weinberg, Phys. Lett. B 62 (1976) 111.
\item B.A. Ovrut and J. Wess, Phys. Rev. D 25 (1982) 409.
\item K. Fujikawa, B.W. Lee and A.I. Sanda, Phys. Rev. D 6 (1972) 2923.
\item  E. Witten, Nucl. Phys. B 185 (1981) 513.
\item  S. Coleman and E. Weinberg, Phys. Rev. D 7 (1973) 888.
\end{enumerate}
\end{document}